\documentclass[aps,prb,twocolumn,showpacs,preprintnumbers,amsmath,amssymb]{revtex4}

\usepackage{graphicx}
\usepackage{dcolumn}
\usepackage{bm}

\begin{document}

\preprint{(\textsc{preprint to PRB})}

\title{Direct Observation and Quantitative Analysis of Anti-ferroelectric Fluctuations in Pb(Mg$_{1/3}$Nb$_{2/3}$)O$_{3}$ Relaxor}


\author{A. Tkachuk}
\email{tkachuk@aps.anl.gov}
\thanks{Current address: XFD division, Advanced Photon Source, Argonne National Laboratory}

\author{Haydn Chen}%
\email{aphchen@cityu.edu.hk}
\thanks{Current address: Department of Physics and Materials Science, City University of Hong Kong, Kowloon, Hong Kong}

\affiliation{%
Department of Materials Science and Engineering and Materials Research Laboratory,
University of Illinois at Urbana-Champaign, Urbana, IL 61801
}%



\date{\today}

\begin{abstract}
Synchrotron x-ray scattering studies of anti-ferroelectric nanoregions (AFR) as fluctuations, which give rise to 1/2(hk0) superlattice reflections ($\alpha$ spots), were systematically performed in Pb(Mg$_{1/3}$Nb$_{2/3}$)O$_{3}$ (PMN) ferroelectric relaxor. Separation of the $\alpha$ spots from underlying diffuse scattering background allowed studying them as separate entities for the first time. AFR fluctuations were shown to be different from the chemical nanodomains (CND) and ferroelectric polar nano-regions (PNR).  Instead, they are formed by anti-parallel short-range correlated Pb$^{2+}$ displacements that lead to unit cell doubling in $\langle110\rangle$ directions, based upon structure factor calculations. Correlation length is only on the 30~\AA \ scale, which defines the effective average size of the AFR fluctuations.  This size remains temperature independent while the total number of AFRs significantly increases near the Vogel-Fulcher freezing temperature of  220~K.  Presence of local fluctuations produced by anti-parallel atomic displacements can explain relaxor behavior as a result of competition between ferroelectric and anti-ferroelectrics type polar ordering in the glass-like systems. 
\end{abstract}


\maketitle

\section{Introduction}
	Pb(Mg$_{1/3}$Nb$_{2/3}$)O$_{3}$ (PMN) belongs to a special class of relaxor ferroelectrics with average cubic perovskite ABO$_{3}$ crystal structure \cite{AVT.51,AVT.52} ($Pm\bar{3}m$), where $A$ sites are occupied by Pb$^{2+}$ and $B$ sites by a mixture of Nb$^{5+}$ and Mg$^{2+}$ ions \cite{AVT.45}. As-grown PMN single crystals exhibit excellent crystalline properties with mosaic spread $\omega \le 0.01^{\circ}$. However, on the nanometer scale there is a significant degree of chemical and displacement disorder \cite{AVT.90, AVT.54, AVT.45, AVT.146, AVT.23, AVT.208, AVT.218,AVT.52, AVT.51,AVT.185,AVT.58, AVT.107, AVT.183, AVT.228, AVT.223,krause71}. Moreover, atoms on the $B$ sites are short-range ordered within kinetically quenched-in chemical nanodomains (CND) with $Fm\bar{3}m$ symmetry \cite{krause71,AVT.90,AVT.23,AVT.54,AVT.146,AVT.208}.  
	
	Strong frequency dispersion of real and imaginary parts of dielectric susceptibility is one of the most recognized properties that distinguish relaxors from "normal ferroelectrics" \cite{AVT.45}. Temperature dependence of the dielectric constant in PMN exhibits a broad peak at $T_{m}\sim$ 265~K with a magnitude exceeding 30,000 when measured under $f=1~kHz$ AC electric field \cite{AVT.45}. Both magnitude and position of the peak strongly depend on frequency, which results in shifts of the dielectric peak maximum towards $T_{f}\sim$220~K as $f$ approaches zero \cite{AVT.49,AVT.45}. 

	Additionally, degree of dispersion, size of CND and structural transitions can be controlled in PMN by various atomic substitutions on both $A$ and $B$ sites. For example, Ti doping leads to formation of $[Pb(Mg_{1/3}Nb_{2/3})O_{3}]_{1-x}-[PbTiO_{3}]_{x}$  (PMN-xPT) solid solutions, where relaxor properties diminish as PT doping increases and eventually  disappear near morphotropic phase boundary corresponding to $x\approx0.34$ on the  PMN-PT phase diagram \cite{AVT.53,AVT.121}.

	Contrarily to  "normal ferroelectrics", in PMN there is no clear crystallographic evidence for any measurable spontaneous macroscopic structural phase transition in a wide (5-800~K) temperature interval \cite{AVT.51, AVT.52}. Instead, it is believed that gradual freezing into nonergodic glass-like state is taking place below $T_{f}$ \cite{AVT.49, AVT.45, AVT.33,AVT.58,AVT.137,AVT.121,AVT.224}. Nevertheless, small rhombohedral distortions \cite{AVT.64} and a change of slope in the temperature dependence of the lattice constant \cite{AVT.223} were also reported near $T_{f}$ from single crystal diffraction studies. It was also demonstrated that macroscopic ferroelectric structural phase transition can be induced in PMN(111) by applying external DC electric field with a magnitude above the critical value of E${_c}\approx$1.7 kV/cm \cite{AVT.126, AVT.21,AVT.33,AVT.224}.

Several models were proposed to explain relaxor behaviors in Pb-containing perovskite compounds \cite{AVT.45,AVT.48,AVT.49,AVT.208,AVT.193,AVT.137}. Current understanding of  PMN relaxor properties relies on existence of polar nanoregions (PNR), possibly with rhombohedral $R\bar{3}m$ symmetry \cite{AVT.218, AVT.52, AVT.51,AVT.67}. These nanoregions are commonly envisioned as interacting ferroelectric fluctuations, which were proposed to nucleate below $T_{d}\approx620~K$ by Burns et~al.~ \cite{AVT.218}. Spherical-Random-Bond-Random-Field (SRBRF) model \cite{AVT.107} is one of the latest  theoretical treatments describing interactions between PNR in a self-consistent fashion. Alternatively, relaxor properties were attributed to special role of Pb atoms causing frustration of the local structure by displacements needed to accommodate lone electron pair of the Pb$^{2+}$ ion in the presence of underlying chemical disorder \cite{AVT.208, AVT.121}.  

Deviation of the local structure from the cubic average is expected due to the presence of aforementioned CND  \cite{AVT.90,AVT.54,krause71,AVT.23,AVT.184,AVT.31,AVT.208,AVT.146} and PNR \cite{AVT.218,AVT.48,AVT.49,AVT.107,AVT.213,AVT.224,AVT.223,hirota02,wakimoto02a} nanoregions.  Because their sizes are believed to be smaller than $100~\AA$,  it is a challenging task to study these nanoregions with inherently long-range order probing tools such as electron, neutron and x-ray diffraction. Nevertheless, it has been demonstrated that chemical short-range order within CND can be directly observed and studied in PMN single crystals by applying  standard crystallographic techniques due to the presence of very weak and diffuse 1/2(hkl) type superlattice peaks, commonly called F spots in literature \cite{krause71, AVT.54, AVT.90,AVT.184, AVT.31}. 

Various diffraction studies of the PMN have also reported the presence of strong diffuse scattering near fundamental Bragg reflections \cite{AVT.58,AVT.54,AVT.51,AVT.52,AVT.67,AVT.25,AVT.111,AVT.223,AVT.228}. Quantitative analysis of temperature dependent neutron diffuse scattering near Bragg peaks has been attributed to polarization fluctuations, which could be related to aforementioned PNRs \cite{AVT.67}. As a result, directions and relative magnitudes of the ionic displacements were determined and ferroelectric nature of parallel $\langle111\rangle$ Pb$^{2+}$ displacements has been established. However, calculated ionic displacement pattern did not have expected center of mass, required by dispalcements originating from optic vibrations. Alternatively, the anisotropic x-ray diffuse scattering near Bragg peaks, which have odd integer sum of the Miller indices, was attributed to pure transverse optic (TO) soft modes, which undergo q-dependent freezing within $\sim$190-260~K temperature range \cite{AVT.25, AVT.111}. Hirota et~al.~claimed to resolve controversy of the diffuse scatterings origin by introducing a phase-shifted soft mode model of PNRs \cite{hirota02}.  Moreover, recent  inelastic neutron scattering experiments \cite{AVT.11, AVT.213} identified presence of the TO soft mode above T$_{d}$, which becomes overdamped at lower temperatures due to condensation of the PNRs \cite{AVT.213}. Interestingly, subsequent recovery of the Curie-Weiss behavior coincides with the freezing temperature T$_{f}$ \cite{wakimoto02a}.  The presence of a distinct thermodynamic phase transition near T$_{f}$ into a nonergodic frozen dipolar glass state was also suggested in the past based on electroacoustic studies of PMN \cite{AVT.137}.

In addition to the F spots, 1/2(hk0) type diffuse superlattice reflections, referred as $\alpha$ spots, have been identified in PMN from the original electron diffraction studies \cite{AVT.54}. These reflections are commonly attributed to unit cell doubling along $\langle110\rangle$ cubic directions caused by chemical or displacement short-range ordering. However, limited number of experimental observations often contradict with each other  in regard to the existence and origins of the $\alpha$ spots \cite{AVT.54,AVT.184,AVT.201,AVT.228, AVT.31,AVT.183}. Interestingly, temperature dependence of the $\alpha$ spots in related Pb(In$_{1/2}$Nb$_{1/2}$)O$_{3}$ (PIN) variable order relaxor was shown to be correlated with long-range macroscopic anti-ferroelectric phase transition in the case of the fully ordered PIN  \cite{AVT.27}.

According to our preliminary  results \cite{AVT.198, tkachuk02a}, ambiguity related to the detection \cite{AVT.31,AVT.184, AVT.183} and temperature dependence of the $\alpha$ spots  \cite{AVT.54, AVT.201, AVT.228} can be resolved by taking into account their extremely weak intensities, broad width  and proper separation from anisotropic diffuse scattering, which appears to extend all the way to the Brillouin zone boundaries occupied by the $\alpha$ spots.  Detailed analysis and discussion of the $\alpha$ spots will be presented  in the following sections. This work is primarily focused on physical origins and quantitative analysis of the $\alpha$ spots in PMN below freezing temperature $T_{f}\approx$220~K.

\section{Experimental}

All scattering studies were performed on PMN single crystals from different sources grown by Czochralski and Bridgeman methods. PMN crystals doped with PbTiO$_{3}$ (PMN-xPT) x\ensuremath{\leq}0.32 were grown by the melted flux method. Crystals were in the form of  $\langle001\rangle$ or $\langle111\rangle$ oriented platelets having surfaces with linear dimensions no larger than 3-7 mm and thickness ${\sim}$1 mm. PMN(111) crystals were sputtered with gold for in-situ x-ray measurements under applied electric field (up to 4 kV/cm). All crystals were of good quality with a mosaic spread no worse than 0.01\ensuremath{^\circ}, obtained from x-ray diffraction rocking curve measurements. Dielectric spectroscopy results obtained from the same crystals were published previously \cite{AVT.114, AVT.175}.

	Synchrotron x-ray work was conducted on X-18A beamline at the National Synchrotron Light Source (NSLS), Brookhaven National Laboratory, and at 33-ID beamline at the Advanced Photon Source (APS), Argonne National Laboratory. Both beamlines used focusing mirrors, which also served as high energy harmonic discriminators. Crystals were studied in 10-300 K range inside closed cycle He gas cryostats mounted on 4-circle or 6-circle kappa diffractometers.

	Energy of the incident x-rays at NSLS was chosen to be 10 keV for optimal beamline performance. At APS the x-ray energy in addition to 10 keV was tuned near Nb ${K}$ (18.99 keV) and Pb $L_{III}$ (13.035 keV). The size of the incident beam was collimated by a pair of slits set to 0.5 $\times$ 0.5 mm$^{2}$ at APS, 1 $\times$ 1 mm$^{2}$ at NSLS. High flux of the synchrotron sources allowed detector slit size no larger than 1~$\times$~1~mm$^{2}$. Typical counting times for diffuse scattering measurements was 1 sec/point at APS and  $\leq$10 sec/point at NSLS. High energy harmonics in the incident x-ray beam for all of our experiments were suppressed to the level that did not cause any observable data contamination. Distribution of the scattered intensity in the reciprocal space was measured by fully automated computer-controlled reciprocal space scans. 2D mesh scans allowed for direct  point-by-point reciprocal space mapping. The interval between the scan points was \ensuremath{\sim}0.01-0.02 reciprocal lattice units (r.l.u.), where 1~r.l.u. is ${2 \pi}/a \approx$1.55~\AA$^{-1}$ and $a\approx$4.04~{\AA} is PMN's lattice constant. 2D diffuse scattering measurements were performed 0.1~r.l.u. away  from the Bragg peak centers.

\section{Results and Discussion}

\begin{figure}
\includegraphics[width=3.4 in]{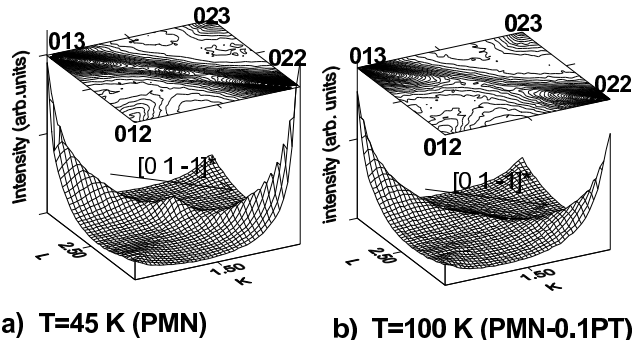}%
 \caption{\label{Fig_1}  Diffuse scattering intensity map measured 0.1 (r.l.u.) away from four $Pm\bar{3}m$ Bragg peaks at: (a) 45 K in PMN. 1/2(035) $\alpha$ spot is in  the middle of the diffuse ridge; (b) 100 K in PMN-0.1PT. Similar diffuse ridge is present but no $\alpha$ spot.}
 \end{figure}

	  Distribution of the diffuse scattering intensity in the KL reciprocal plane defined by the (012), (013), (023) and (022) Bragg reflections at four corners is shown in Fig.~\ref{Fig_1} for PMN (at 45~K) and PMN-0.1PT (100~K).  Horizontal axes (x and y) correspond to K ($[010]^*$) and L ($[001]^*$) reciprocal lattice directions with  intensity plotted along the vertical z axis on the linear scale. In addition, contour plots are displayed above the corresponding surface plots  in Fig.~\ref{Fig_1} for better visualization. Only the diffuse scattering tails measured $\sim0.1$ (r.l.u.) away from the Bragg peak maxima are visible in Fig.~\ref{Fig_1}.  Centers of the Bragg peaks  are outside the plot boundaries because  their intensities are more than $10^8$ times greater. 
	  
	Two pronounced $[01\bar{1}]^*$ diffuse scattering ridges connecting (013) and (022) Bragg peaks can be readily identified on both plots in Fig.~\ref{Fig_1}.    Fig.~\ref{Fig_1}(a) also demonstrates the existence of a very weak and diffuse 1/2(035) $\alpha$ spot located half way between (013) and (022) Bragg peaks in PMN at 45~K. The FWHM of this peak along $[01\bar{1}]^*$ direction corresponds to $\sim$0.1 (r.l.u.), which is  more than 100 times larger than the width of the fundamental Bragg peaks.  Due to the inverse relationship between the reflection width and the correlation radius in the real space defined by the well-known Scherrer relationship \cite{warren69}, the corresponding length scale of the $\alpha$ spot is about 30 $\AA$, which is  comparable in size with previously mentioned CND ($\sim$50~$\AA$).  Moreover, there is no clear evidence for such a peak for data in Fig.~\ref{Fig_1}(b),  obtained with identical  experimental setup. Therefore, absence of the 1/2(035) $\alpha$ spot in Fig.~\ref{Fig_1}(b) together with the FWHM argument excludes the origin of the $\alpha$ superlattice peak in PMN from diffraction of the $\lambda/2$ contaminating x-ray harmonic. 
	
	Previous reports of the F and $\alpha$ spots measured with x-rays were presented from the linear 1D reciprocal scans  \cite{AVT.184,AVT.31,AVT.201}.  Also, transmission electron microscopy (TEM) images (2D reciprocal maps) clearly show $\langle0\bar{1}1\rangle$ diffuse ridges similar to the one in Fig.~\ref{Fig_1}, which intersect locations of the 1/2(hk0) spots \cite{AVT.54, AVT.228}, e.~i.~ $\alpha$ spots in our notation. Therefore, if these diffuse scattering ridges are 1D linear features in the reciprocal space, any scan that intersects their line of direction  will produce a peak cross section regardless of  existence of the 1/2(035) peak in Fig.~\ref{Fig_1}a. This means that if the $\alpha$ spot is a real peak produced by 3D correlations, it must exhibit peak cross section in any direction regardless of the direction of the reciprocal scan. 
	
\begin{figure}
\includegraphics[width=3.4in]{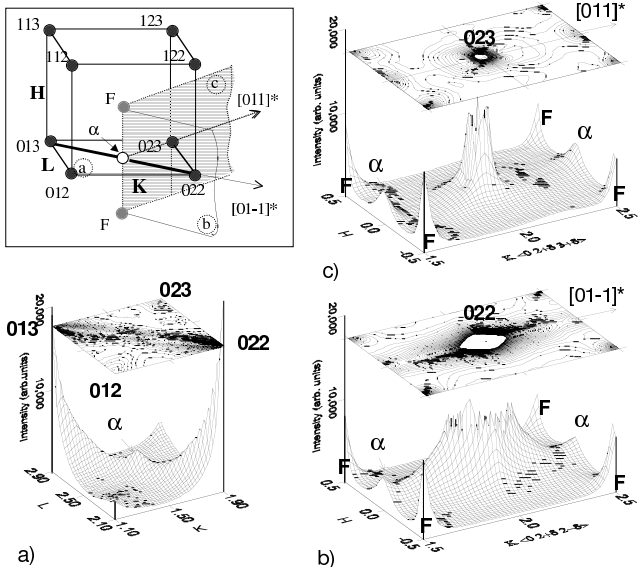}%
 \caption{\label{Fig_2} Intensity maps measured $\sim$0.1 (r.l.u.) away from (022) Bragg peak in three mutually perpendicular reciprocal space planes: (a) T=45 K; (b) and (c) T=160 K. Presented data proves that diffuse ridge in (a) is due to diffuse scattering concentrated primarily along $[01\bar{1}]$*.}
 \end{figure}

 	 In order to prove this point, we measured diffuse scattering intensity in three mutually perpendicular reciprocal planes that contain 1/2(035) $\alpha$ spot  from Fig.~\ref{Fig_1}(a).  Experimental data is presented in Fig.~\ref{Fig_2} for corresponding $a$, $b$ and $c$ planes. The locations of these three planes relative to the reciprocal unit cell can be identified from the sketch depicted on the top left corner of the figure. Cube corners marked with black circles correspond to fundamental cubic perovskite Bragg reflections. F spots and $\alpha$ spots can be identified on  body-centered (grey circles) and face-centered (white circle) cubic positions, respectively. Note that only one 1/2(035) $\alpha$ spot is shown to reduce clutter. Fig.~\ref{Fig_2}(b) and (c) also indicate the absence of edge-centered peaks, called $\beta$ spots in related variable order $Pb(In_{1/2}Nb_{1/2})O_{3}$ (PIN) relaxor \cite{AVT.27}, which once again proves the absence of any detectable $\lambda/2$ contamination in the incident x-ray beam. Data in Fig.~\ref{Fig_2}(a) are the same data shown in Fig.~\ref{Fig_1}(a), while  data in Fig.~\ref{Fig_2}(b) and (c) were taken at 160 K and plotted on the same vertical scale for direct intensity comparison. Both planes $a$ and $b$ contain $[01\bar{1}]^*$ direction, which is drawn in the sketch with a thick line connecting (013) and (022) Bragg peaks. Observation of the 1/2(035) $\alpha$ spot  along any reciprocal  direction in these planes unambiguously  proves its 3D origin. It is important to note from Fig.~\ref{Fig_2}(b) and (c), that less than 50~\%~of the  total peak cross section at H=0 along the H axis ( e.~i.~[100]* direction), corresponds to the actual $\alpha$ spot superlattice peak, assuming that diffuse ridge is from different origins. In fact, the first report of the  ${\alpha}$ spot with synchrotron x-rays was also presented using  linear [100]* reciprocal scan, which may be contaminated by the $[01\bar{1}]$* diffuse ridge \cite{AVT.184}.

	Moreover, if the cross section of the diffuse ridge had not been accounted for,  the presence of the $\alpha$ spots  would have been concluded for PMN-0.1PT at 100 K, which is in fact not the case according to Fig.~\ref{Fig_1}(b).  Our measurements in a wide 15-800 K temperature interval have shown that peak cross sections of the ridges near zone boundary do not change even above the Burn's temperature $T{_d}$ \cite{tkachuk02a}. Since diffuse ridges in the form of 1D linear features are always present, observation of the $\alpha$ spots above $\approx$220~K  by other workers can be explained by measuring primarily diffuse ridge cross sections instead of actual $\alpha$ spots \cite{AVT.228,AVT.54,AVT.184}. 
	
	On the other hand, it is well known that diffuse scattering in PMN at $q\ll$0.1 (r.l.u.) from the Brillouin zone centers exhibits strong temperature dependence \cite{AVT.184,hirota02,AVT.25}. In contrast, temperature independent diffuse scattering at zone boundaries (q=0.5~r.l.u.) suggests that diffuse scattering observed in different parts of the reciprocal space comes from different origins. For example, temperature independent $\langle1\bar{1}0\rangle$* diffuse scattering ridges were reported to exist in related Pb(Sc$_{1/2}$Nb$_{1/2}$)O$_{3}$ (PSN) relaxor \cite{AVT.28}. The origin of the ridges has been attributed to static or dynamic displacements in $\{1\bar{1}0\}$ planes without (or with weak) correlation between the planes along  $\langle1\bar{1}0\rangle$ directions. Observation of the $\alpha$ spots  in PSN with synchrotron x-rays was explained by Takesue et~al.~due to existence of the linear anti-ferroelectric chains  \cite{AVT.5}. However, displacement correlation lengths were calculated from the widths of the peaks measured at the positions of the $\alpha$ spots by three $\langle100\rangle$* scans without taking into account cross sections of the overlapping $\langle0\bar{1}1\rangle$ diffuse ridges. From now on, if not otherwise mentioned, all the quantitative  $\alpha$ spot analyses have been performed along the direction of the $\langle01\bar{1}\rangle$* diffuse ridges.
	
	Glazer studied the origins of the different superlattice reflections in perovskites \cite{AVT.95}. According to his proposed classification scheme, the origin of the $\alpha$ spots can be attributed to correlated in-phase oxygen octahedra rotations about the axis defined by even index in the 1/2(hkl) notation \cite{AVT.95}.  In this case, the presence of the 1/2(035) $\alpha$ spot in Fig.~\ref{Fig_2} would correspond to unit cell doubling along $\langle011\rangle$ directions, which should lead to local tetragonal distortions along the rotation axis $a$. Note that correlation length of these in-phase rotations must be $\sim$30~\AA~ according to the width of the $\alpha$ spot.

	It has been also shown that Pb$^{2+}$ ions are significantly displaced by about 0.3~$\AA$ from their equilibrium positions according to pair distribution function (PDF) measurements in PMN \cite{AVT.208}. However, PDF measurements cannot answer the question along which particular directions these Pb ions are displaced, due to local probe nature of this experimental technique. On the other hand, Bragg peak structure factor refinements \cite{AVT.51,AVT.52,AVT.185} and analysis of the polarization vectors from neutron diffuse scattering \cite{AVT.67,hirota02} in PMN have inferred existence of nonuniform parallel Pb displacements preferential along $\langle110\rangle$ or rhombohedral $\langle111\rangle$ polar directions at low temperatures.  However, in order to contribute to the structure factor of the $\alpha$ spots, some fraction of the Pb atoms must be also short-range correlated in $\langle011\rangle$ directions in an anti-parallel fashion, required for unit cell doubling. Indeed, our structure factor calculations, which will be discussed later in this paper, prove direct involvement of the Pb displacements in production of the 1/2(hk0) superlattice reflections. 

\begin{figure}
\includegraphics[width=3.4in]{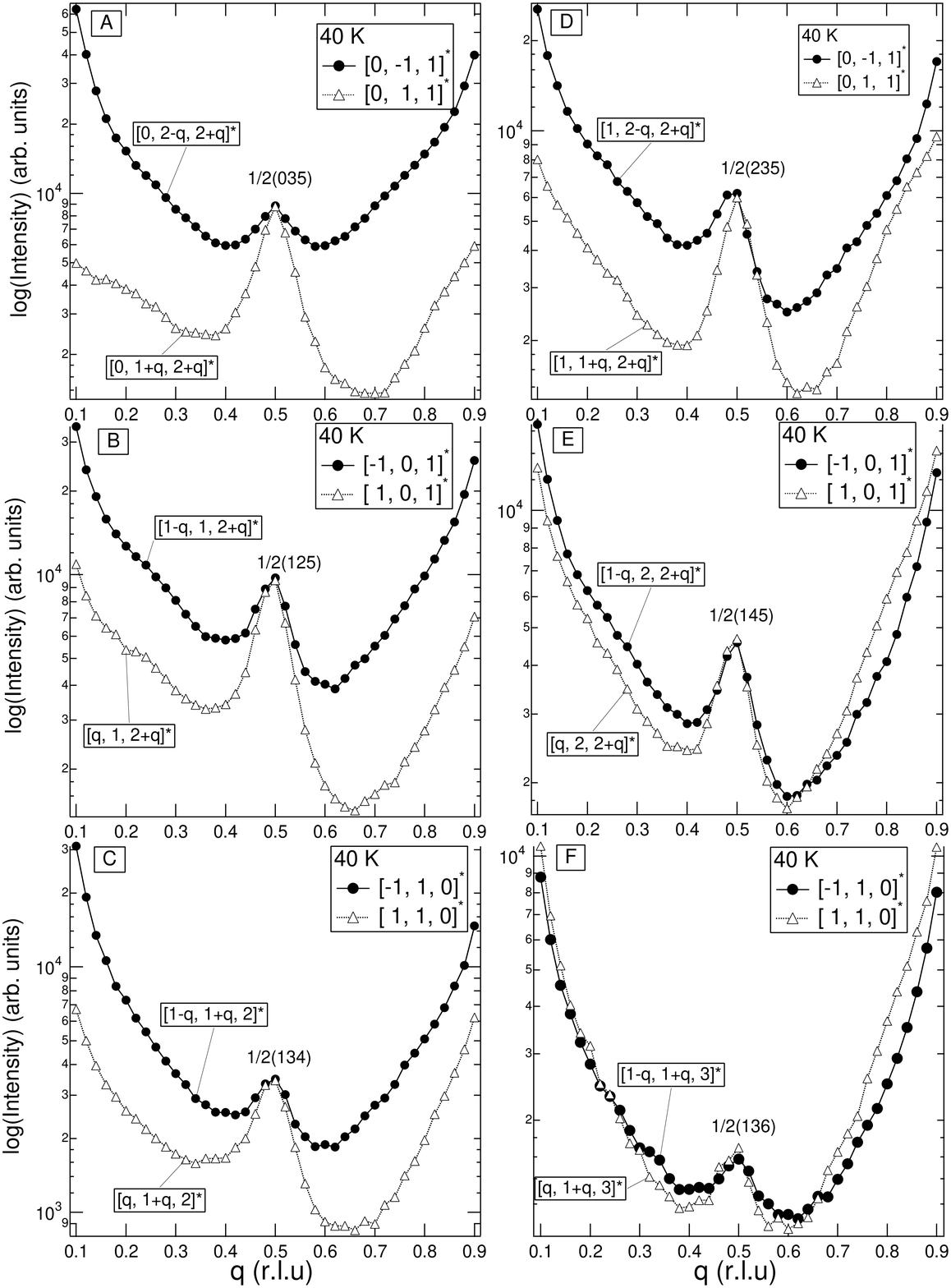}%
 \caption{\label{Fig_3} Reciprocal cubic face diagonal scans in the reciprocal unit cell shown in Fig.~\ref{Fig_2}. Scans along $\langle01\bar{1}\rangle^*$ (solid circles) and the scans along $\langle011\rangle^*$ (empty triangles).}
\end{figure}

 	Since average PMN structure is cubic at all temperatures, displacements that give rise to  $\alpha$ spots must be correlated on the nanometer scale along all symmetry equivalent cubic $\langle011\rangle$ directions, which would lead to appearance of  the $\alpha$ spots on every face-centered position in the reciprocal unit cell.  Indeed, we found $\alpha$ spots on all six reciprocal cubic faces shown Fig.~\ref{Fig_2} sketch below $220~K$ in PMN.   Corresponding linear scans along the  direction of the diffuse scattering ridge ($\langle01\bar{1}\rangle^*$) and scans perpendicular to this direction ($\langle011\rangle^*$) are shown in Fig.~\ref{Fig_3}. Scans along $\langle01\bar{1}\rangle^*$ are plotted using solid circles and the scans along $\langle011\rangle^*$ directions  are plotted with empty triangles. Peaks on the solid circle  curves at q=0.5 are the actual $\alpha$ spots "sitting" on top of the large background created by the $\langle01\bar{1}\rangle^*$ diffuse ridges. Peaks on the empty triangle curves are produced by the contributions from both $\alpha$ spots and the cross sections of the diffuse ridges. These data clearly demonstrate that large errors would occur if integrated intensities of the ${\alpha}$ spots are extracted from the empty triangle scans rather than from the solid circle ones. The relative intensities of the $\langle01\bar{1}\rangle$* ridges are different and appear to be correlated with the structure factors of the Bragg peaks that they connect. For example, 1/2(145) and 1/2(136) $\alpha$ spots in Fig.~\ref{Fig_3}(e) and (f), which are located between relatively weak Bragg peaks, are the least affected by the anisotropy of the diffuse scattering background.
	
\begin{figure}
\includegraphics[width=3.4in]{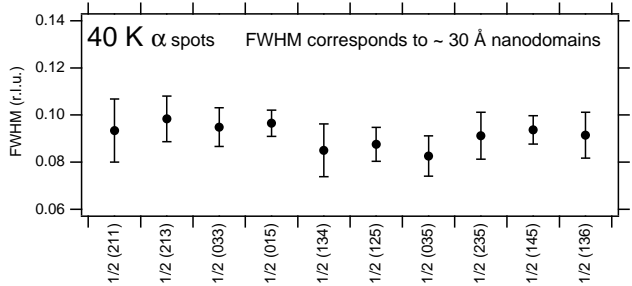}%
 \caption{\label{Fig_4}  FWHM of various $\alpha$ spots separated from the background along $\langle01\bar{1}\rangle$* diffuse scattering ridge directions.}
\end{figure}

	Extracted FWHM values for several $\alpha$  spots, including solid circle scans in Fig.~\ref{Fig_3} , are shown in  Fig.~\ref{Fig_4}. The average size of the fluctuations responsible for the $\alpha$ spots corresponds to $\sim30~\AA$ from Scherrer equation. Notice that FWHM is the same, within experimental errors,  for all the reflections in  Fig.~\ref{Fig_4}  independent of the reflection's distance from the origin of the reciprocal space. This indicates that the width is predominantly defined by the size of the nanoregions rather than by strain effect, which typically increases the width of the diffraction peaks located further away from the center of the reciprocal space.
	 
\begin{figure}
	\includegraphics[width=3.4in]{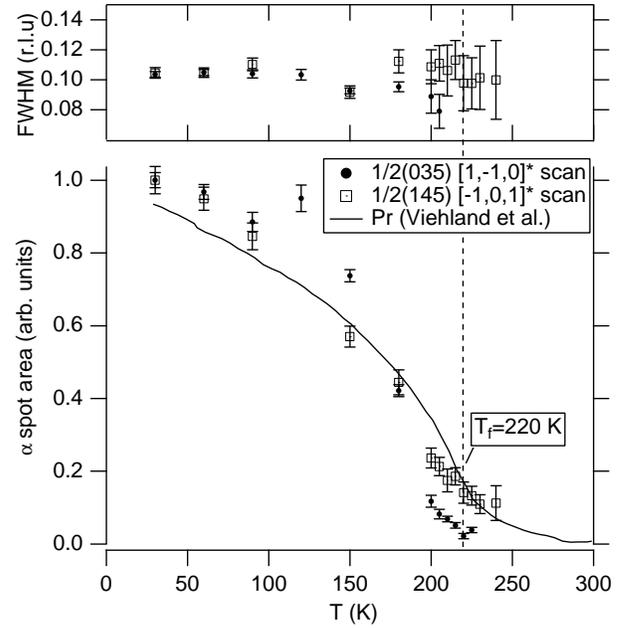}%
 \caption{\label{Fig_5}  Temperature dependence of the $\alpha$ spot's FWHM and integrated intensity vs.~temperature. Direct evidence of local structure changes near freezing temperature T$_{f}\approx$220~K. Normalized remanent polarization P${_r}$ is from Ref.\onlinecite{AVT.190} as a guide to the eye.}
\end{figure}

	Figure~\ref{Fig_5} presents temperature dependence of the integrated intensity and FWHM values of the $\alpha$ spots measured along $\langle01\bar{1}\rangle$* diffuse ridge without electric field.  Normalized remanent polarization P$_{r}$ digitized from Ref. \cite{AVT.190} is also shown on the same graph as a guide to the eye. Remarkably, significant reduction in the intensity of the $\alpha$ spots on heating occurs near the phenomenological freezing phase transition T$_{f}$, which is not accompanied by any kind of Bragg peak splitting \cite{AVT.52,AVT.51} in the absence of applied electric field \cite{AVT.126,AVT.21}. This does not contradict average cubic structure, since FWHM$\approx$0.1 (r.l.u.) of the $\alpha$ spots corresponds to ${\sim}$30 {\AA} correlation length. It also appears to be temperature independent below T$_{f}$ within experimental errors. 

\begin{figure}
\includegraphics[width=3.0in]{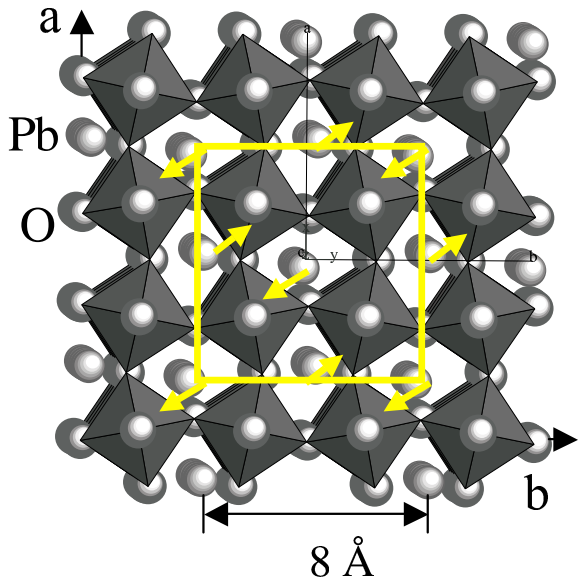}%
 \caption{\label{Fig_6} Displacement pattern, which doubles unit cell along one of the equivalent  $\langle110\rangle$, e.~i. $[110]$. Directions of the anti-parallel Pb displacements are indicated with arrows.}
\end{figure}

Integrated intensity $I$ of the diffraction peaks is proportional to their structure factor F ($I \sim |F|^{2}$), which can be directly calculated if the positions of all  atoms in the unit cell are known \cite{warren69}.
According to earlier discussion, integrated intensities of the $\alpha$ spots must be extracted from the scans along the diffuse ridge directions (e.~g. solid circles in Fig.~\ref{Fig_3}). In order to calculate structure factors of the $\alpha$ spots, we considered a doubled unit cell ($a\approx 8~\AA$). Figure~\ref{Fig_6} shows a proposed displacement pattern where anti-parallel Pb displacements double unit cell along $[1\bar{1}0]$ direction. In-phase oxygen octahedra rotations along the $c$ axis (perpendicular to the plane of the graph) are also shown as another possible contribution to the structure factor of the $\alpha$ spots. Corners of the the octahedra are oxygen atoms, Nb/Mg are located inside of the oxygen ocathedrons and Pb atoms, displaced $0.3~\AA$ from the equilibrium sites, are positioned interstitially in the oxygen octahedra network. 

\begin{figure}
\includegraphics[width=3.4in]{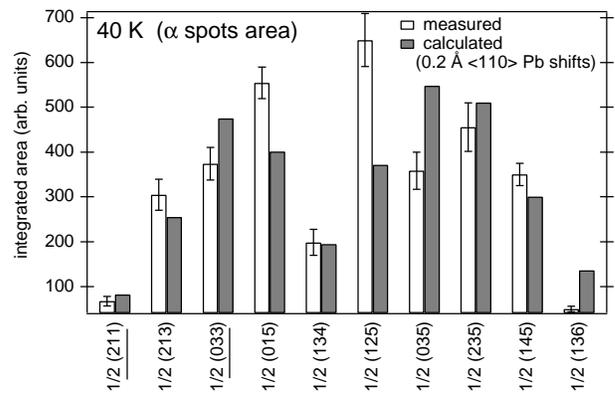}%
\caption{\label{Fig_7} Observed and calculated $\alpha$ spot  integrated intensity for  $0.2~\AA$  $\langle110\rangle$ anti-parallel Pb displacements correlated in 30 $\AA$ nanodomains.}
\end{figure}

	Measured at 40 K, integrated intensities of the  $\alpha$ spots are represented as white bars in Fig.~\ref{Fig_7} after performing standard absorption and Lorentz factor data corrections \cite{warren69}.  Structure factor modeling without Pb displacements (only pure oxygen octahedra rotations) shows that $\alpha$ spots with identical odd indices are forbidden \cite{AVT.95,tkachuk02b}.  However, reflections, such as  1/2(211) and 1/2(033), were experimentally observed and their corresponding integrated intensities are nonzero according to Fig.~\ref{Fig_7} .  The results of the least squares fit with only two variable parameters are presented as gray bars. All reflections were fitted at once. One of the fitting parameters was a magnitude of the Pb displacements \ensuremath{\delta}, and another was common for all peaks intensity scaling factor. Debye-Waller factors were taken from the literature \cite{AVT.67,AVT.185,AVT.183}. Structure factor calculations showed, that pattern in Fig.~\ref{Fig_6} would only contribute to the $\alpha$ spots, which have even third Miller index ($l$), such as 1/2(134) and 1/2(136) in  Fig.~\ref{Fig_3}. \ Similar patterns with Pb displacements having 30 {\AA} correlation length along other equivalent $\langle011\rangle$ directions will contribute to reflections with $h$ or $k$ even indices, accordingly.  
	
	The actual displacement pattern may be more complicated than the one presented in  Figure~\ref{Fig_6}, which could contribute to the differences between observed and calculated structure factor results for some reflections in Fig.~\ref{Fig_7}. Inclusion of Nb displacements or in-phase oxygen octahedra rotations with angles up to 10$^{\circ}$ only marginally improved the fit. Neutron measurements, which are more sensitive to oxygen, are clearly needed in order to better understand the role of oxygen octahedra rotations. What is important is that oxygen octahedra alone cannot explain $\alpha$ spot structure factor without Pb displacements. Contribution of Pb displacements to the $\alpha$ spots is consistent with differential anomalous factor scattering (DAFS) measurements near Pb L$_{III}$ absorption edge \cite{tkachuk02b}. On the contrary, we were not able to find any evidence for Nb contribution from the DAFS measurement near Nb $K$ edge \cite{tkachuk02b}.
	 
	Interesting conclusions can be made from a comparison of the $\alpha$ and F spots in PMN with the superlattice reflections found in PMN-0.06PT single crystal \cite{AVT.198}.  According to our measurements, the sizes of the nanoregions that give rise to the $\alpha$ spots in PMN-0.06PT and pure PMN are identical within experimental errors, i.~e. $\approx$30~$\AA$. At the same time, the size of the CND regions in PMN-0.06PT, obtained from the width of the F spots, is only $\approx$25 $\AA$, which is half the size of CND regions in PMN.  These results together with observed strong temperature dependence of the $\alpha$ spots \cite{AVT.198}, indicate that their origins must be different and unrelated to the CND regions where short-range chemical ordering is quenched and therefore temperature invariant \cite{AVT.146}.

	We have shown that $\alpha$ spots are distinct 3D peaks, which means that producing them nanoregions scatter radiation coherently. If they scattered radiation incoherently, we would have observed  "powder rings" in the reciprocal space, as in the case of powder diffraction \cite{warren69}.  Therefore, intensity of the $\alpha$ spots is determined either by the total number of these nanodomains or by  magnitude of the atomic displacements.  	In a  separate experiment we observed expected \cite{AVT.126,AVT.21} rhombohedral splitting of the Bragg peaks in PMN(111) \cite{tkachuk02b} under electric field $\ge$1.8 kV/cm. However, we did not register any changes in either FWHM or intensity of the \ensuremath{\alpha} spots in both field cooled and zero field cooled regimes below T$_{f}$, in contrast to PNRs \cite{AVT.224}. This fact indicates that changes in the average macrostructure have no effect on either size or number of the nanoregions that give rise to the \ensuremath{\alpha} spots. 

It is not possible to determine how many different  $\langle110\rangle$ directions are operative within a single nanodomain.  Important fact is that this type of correlated anti-parallel Pb distortions is expected to compete with parallel ferroelectric Pb displacements believed to be present in PNRs \cite{AVT.52,AVT.67,hirota02}. These new nanodomains, which give rise to $\alpha$ spots, fit the description of anti-polar nanoregions (AFR) due to anti-ferroelectric like displacements of the Pb atoms from one unit cell to another along equivalent $\langle110\rangle$  directions. Absence of $\alpha$ spots in PMN-0.1PT can be due to much smaller total number of anti-ferroelectric nanodomains when compared to pure PMN or PMN-0.6PT, which would be responsible for peaks not resolvable even with synchrotron radiation.  Alternatively, relaxor behavior in PMN-0.1PT  can be explained due to different competing mechanism, such as local $[100]$ Pb displacements without any unit cell doubling \cite{AVT.223}.

\section{conclusions}

   Anti-ferroelectric ordering is established by correlated atomic displacements which produce dipole moments with anti-parallel arrangement throughout the crystal.  Studies of anti-ferroelectric nanodomains  with techniques sensitive to electric dipoles are challenging, since the net polarization for each nanoregion is zero. 
   
    From diffraction point of view, each type of fluctuations contributes intensity to the different parts of the reciprocal space.  Diffuse scattering from ferroelectric fluctuations (PNR) is mainly concentrated near fundamental Bragg reflections while anti-ferroelectric fluctuations produce diffuse superlattice reflections ($\alpha$ spots in the case of PMN) at the Brillouin zone boundaries as a result of the unit cell doubling. Structural differences between anti-ferroelectric nanoregions and cubic host lattice were studied in this work based on temperature dependence and structure factor modeling of $\alpha$ spots. These peaks are extremely weak, however utilization of third generation synchrotron x-ray sources  allowed us to study their temperature dependence after separation from anisotropic linear $\langle0\bar{1}1\rangle^{*}$ diffuse ridges, which become temperature independent for q$\gg$0.1~(r.l.u.) away from the Brillouin zone centers \cite{tkachuk02a}

Correlation length, obtained from the width of the $\alpha$ spots, is only $\sim$30~{\AA}, which defines the average size of producing these peaks nanodomains. These nanodomains are formed by short-range correlated anti-parallel Pb displacements along equivalent $\langle110\rangle$  directions with a magnitude of \ensuremath{\sim}0.2 {\AA} based upon the structure factor calculations. 
Fluctuations created by these locally correlated displacements are different from chemical nanodomains (CND) and ferroelectric polar nanoregions (PNR); they constitute a new type of fluctuations with anti-ferroelectric type displacement ordering (AFR) based on anti-parallel nature of the Pb displacements.

Freezing phase transition has been identified in PMN near T$_{f}\approx$220 K from the temperature dependence of the integrated intensity of the $\alpha$ spots without application of the external electric field. Significant enhancement in the intensity of the $\alpha$ spots below T$_{f}$ we attribute to increase in a total number of AFRs, which average size ($\sim$30 {\AA}) remains constant down to the lowest measured temperature of 10 K. Nothing can be said about dynamics of these fluctuations from our studies, since interaction time between electrons and x-rays during the scattering process is $\sim$10$^{-15}$~sec. Similar increase in the total number of ferroelectric fluctuations (PNR) near $T_{f}$ was reported from recent macroscopic polarization measurements under applied DC electric field \cite{AVT.224}. Competition between randomly occurring anti-ferroelectric and ferroelectric fluctuations may be responsible for the relaxor behavior in PMN.

\section{acknowledgments}

We would like to thank: Dr. Zschack from the UNICAT ID-33 beamline at the APS, ANL;  Dr. Ehrich from the MATRIX X-18A beamline at the NSLS, BNL;  Dr. Vigliante, Dr. Wohner,  Dr.Kasper and Prof. Dosch from Max Plank Institute, Stuttgart, Germany for technical assistance during the experiments and enlightening discussions.  

Authors also would like to thank Dr.  Colla and Prof. Feigelson  for providing good quality PMN single crystals and stimulating discussions. Special thanks to Dr. Vakhrushev for a very fruitful discussion regarding data interpretation in the final stage of this manuscript preparation.

This research is based upon work supported by the U.S. Department of energy, Division of Materials Sciences under award No. DEFG02-96ER45439 and by the state of Illinois IBHE HECA NWU A207 grant, though the Frederick Seitz Materials Research Laboratory at the University of Illinois at Urbana-Champaign.

National Synchrotron Light Source (NSLS) is supported by the U.S. Departrment of Energy under Contract No. DE-AC02-76CH00016. Use of the Advanced Photon Source was supported by the U.S. Department of Energy, Basic Energy Sciences, Office of  Science, under Contract No. W-31-109-Eng-38.

\bibliography{prb_paper14}

\end{document}